# CNN-Based Invertible Wavelet Scattering for the Investigation of Diffusion Properties of the In Vivo Human Heart in Diffusion Tensor Imaging

Zeyu Deng[a,1], Lihui Wang[a,*], Zixiang Kuai[b,**], Qijian Chen[a], Xinyu Cheng[a], Feng Yang[c], Jie Yang[d], Yuemin Zhu[e]

[a] Key Laboratory of Intelligent Medical Image Analysis and Precise Diagnosis of Guizhou Province, College of Computer Science and Technology, Guizhou University, Guiyang, 550025, China
[b] Imaging Center, Harbin Medical University Cancer Hospital, Harbin, 150081, China
[c] School of Computer and Information Technology, Beijing Jiaotong University, Beijing, 100044, China
[d] Institute of Image Processing and Pattern Recognition, Shanghai Jiao Tong University, Shanghai, 200240, China
[e] University Lyon, INSA Lyon, CNRS, Inserm, IRP Metislab CREATIS UMR5220, U1206, F-69621 Lyon, France

**Abstract**

In vivo diffusion tensor imaging (DTI) is a promising technique to investigate noninvasively the fiber structures of the *in vivo* human heart. However, signal loss due to motions remains a persistent problem in *in vivo* cardiac DTI. We propose a novel motion-compensation method for investigating in vivo myocardium structures in DTI with free-breathing acquisitions. The method is based on an invertible Wavelet Scattering achieved by means of Convolutional Neural Network (WSCNN). It consists of first extracting translation-invariant wavelet scattering features from DW images acquired at different trigger delays and then mapping the fused scattering features into motion-compensated spatial DW images by performing an inverse wavelet scattering transform achieved using CNN. The results on both simulated and acquired in vivo cardiac DW images showed that the proposed WSCNN method effectively compensates for motion-induced signal loss and produces in vivo cardiac DW images with better quality and more coherent fiber structures with respect to existing methods, which makes it an interesting method for measuring correctly the diffusion properties of the in vivo human heart in DTI under free breathing.

## 1. Introduction

Cardiovascular disease remains the world's leading cause of death. More than 17 million people die of cardiovascular disease each year, which represents about 31% of all global deaths [1]. The knowledge about myocardial fiber structure and its link to heart function is fundamental for the diagnosis and the treatment of cardiovascular disease.

Diffusion magnetic resonance imaging (dMRI), especially diffusion tensor imaging (DTI) [2]–[11], is currently the only technique that is increasingly used for investigating noninvasively the myocardial architectures of both normal and diseased hearts. It has been demonstrated that the alternation in myocardial structures or cardiac fiber orientations is highly related to the change of diffusion metrics which are detected in DTI. For instance, fiber orientation disarray [5] and decreased fractional anisotropy (FA) were found in the myocardial infarction region. In hypertrophic cardiomyopathy, FA was decreased and mean diffusivity (MD) was increased [12]. In accurate ischemia, it is usually accompanied by a significant rise in MD and fall in FA [13]. These findings showed that DTI is a promising technique that may provide meaningful imaging biomarkers for early cardiac disease diagnosis.

However, currently, most of cardiac DTI researches were focused on *ex vivo* hearts since DTI is very sensitive to motions, in particular in the case of free breathing. Respiratory and/or cardiac motions can induce important signal loss in diffusion-weighted (DW) images, thus making it difficult to correctly measure diffusion properties and fiber structures of an *in vivo* heart. Therefore, reducing the influence of motions on *in vivo* cardiac DW images is of primordial importance.

To reduce the effects of motions, respiratory gated [14][15], breath-holding [15][16][17] or free-breathing acquisition [18] [19] techniques were used. A number of new acquisition sequences for *in vivo* hearts were also proposed, such as the dual-gated simulated echo sequence [20][21][22], the bipolar PGSE sequence with velocity compensation [23][24], and the segmented balanced steady-state free precession (bSSFP) sequence or turbo spin echo (TSE) sequence with acceleration compensation [9]. Despite these advanced acquisition strategies, removing motion effects in *in vivo* cardiac DW images is still an important issue and a challenging work in cardiac DTI. Alternatively, postprocessing methods were reported in the literature. The postprocessing PCATMIP method was proposed to correct motion effects in cardiac DW images [25], which integrates the principle component analysis (PCA) and temporal maximum intensity projection (TMIP) techniques to recover signal loss caused by heart motions. Inspired by this work, Wei et al. presented a motion correction method called WIF for *in vivo* cardiac DTI based on wavelet fusion [18]. These studies demonstrated that investigating effective post-processing methods makes it possible to compensate for motion effects in *in vivo* cardiac DTI with only clinically available imaging sequences.

In this paper, we propose an invertible wavelet scattering method based on convolutional neural network (CNN) to reduce motion-induced signal loss in DW images of the *in vivo* human heart. The method, called WSCNN (mainly based on Wavelet Scattering and CNN), consists in extracting the features of in vivo cardiac DW images using wavelet scattering, fusing the image feature maps corresponding to different time points, and performing inverse wavelet scattering transform via CNN to recover motion-induced signal loss in *in vivo* cardiac DW images. The proposed method is evaluated on both simulated and acquired *in vivo* cardiac DW images.

## 2. Materials and methods

The overall workflow of the proposed WSCNN method is depicted in Figure 1. Firstly, the DW images acquired with 10 different trigger delays (TDs) are pre-processed to extract the myocardium as ROI. Then, the feature maps of the 10 TDs DW images are extracted using wavelet scattering. After that, the feature maps are fused based on a certain rule. Finally, a deep CNN-based model is explored to learn the relationship between the scattering feature maps and the corresponding DW images using the well-trained model, thus generating the fused DW image from the fused scattering feature maps.

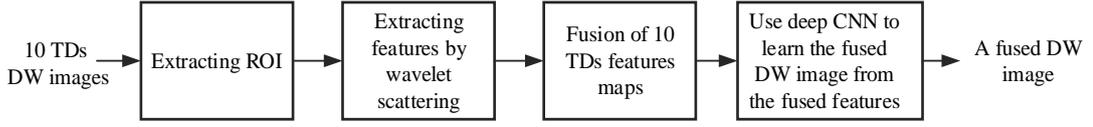

Figure 1. Overall workflow of the proposed WSCNN method.

In the following subsections, the extraction of wavelet scattering features, the fusion of feature maps and the learning of the fused DW images will be described in detail.

**2.1 Wavelet scattering for image feature extraction**

Wavelet transform is a common method for extracting image features via the following transformation

$$Wx = \begin{bmatrix} x * \varphi_j(u) \\ x * \psi_{j,r}(u) \end{bmatrix}, \quad (1)$$

where $x * \varphi_j(u)$ represents the low-frequency information of image $x$ at the scale $J$, and $x * \psi_{j,r}(u)$ the high-frequency information at the scale $j$ and in the direction $r$. $\varphi_j(u)$ is a scaling function expressed as $\varphi_j(u) = 2^{-2j}\varphi(2^{-j}u)$, and $\psi_{j,r}(u)$ is a directional wavelet function expressed as $\psi_{j,r}(u) = 2^{-2j}\psi_r(2^{-j}u)$. Although the wavelet transform can restore the details of the image, it allows obtaining only high-frequency components in three directions (vertical, diagonal, longitudinal) and does not have translation invariance due to convolution operation of wavelet. To deal with these issues, wavelet scattering was proposed by Mallat [26] which allows us to extract either deformation invariant or translation invariant features in multiple directions at multiple scales by introducing a nonlinear operation, namely the modulus of high-frequency coefficients $|x * \psi_{j,r}(u)|$. With such operation, the translation invariant features can be obtained by multiplying a low-pass filtering $\varphi_J(u)$

$$|x * \psi_{j,r}(u)| * \varphi_J(u). \quad (2)$$

Note that the translation invariant features are obtained at the cost of losing high-frequency information. To recover high-frequency information, wavelet decomposition at larger scale is performed on the modulus of the existing high-frequency coefficient, which can be formulated as

$$|x * \psi_{j,r}(u)| * \psi_{j+1,r}(u). \quad (3)$$

However, the feature extracted in this way still lacks translation invariance, so it is necessary to continue the modulus operation and low-pass filtering to achieve the stability of the feature coefficients, such that

$$\left\| x * \psi_{j,r}(u) \right| * \psi_{j+1,r}(u) \right| * \varphi_J(u). \quad (4)$$

From formula (3) and (4), it can be seen that the high-frequency information of each direction is transmitted to the next scale by the modulus operation. Further iteration of wavelet transform and modulus operation can yield more translation invariants. This process, called wavelet scattering, can be expressed as:

$$Wx = \begin{bmatrix} U_m \\ S_m \end{bmatrix}, \quad (5)$$

where the subscript $m$ designates the wavelet scattering level that indicates the number of decomposition operations on the high frequency information, $U_m$ represents the scattering propagation operator

$$\begin{aligned} U_0 &= x & m = 0 \\ U_m &= \left\| x * \psi_{j_1, r}(u) \right| \cdots * \psi_{j_m, r}(u) \right| & m \geq 1 \end{aligned}, \quad (6)$$

and $S_m$ denotes the scattering coefficients expressed by

$$S_m = U_m * \varphi_J. \quad (7)$$

In the present study, wavelet scattering being used to extract translation invariant and deformation invariant features of DW images, we chose the following parameters: wavelet scattering level $m=1$, number of scattering directions at each scale $L=10$, number of wavelet decomposition scales $J=2$. Figure 2 illustrates the process of wavelet scattering, where $x$ is the input image. The blue arrows represent the 0th level of scattering, the green arrows the 1st level of scattering and the red arrows the scattering output at the largest scale $J$, which has translation invariance and deformation invariance.

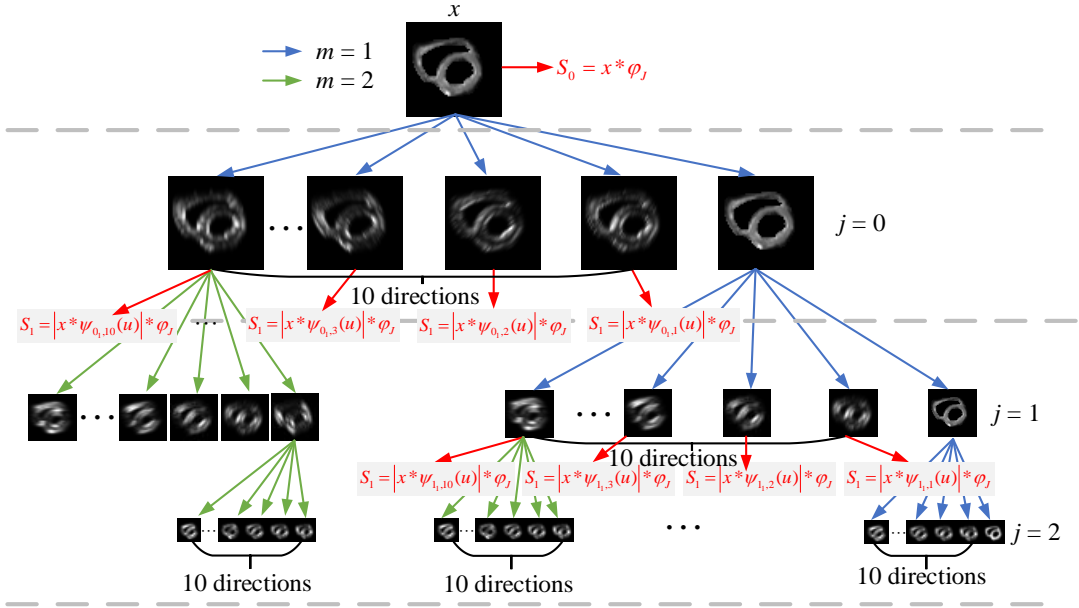

Figure 2. The process of wavelet scattering. Blue arrows: 0th level of scattering. Green arrows: 1st level of scattering. Red arrows: output at the largest scale $J$.

When $m=0$, wavelet transform outputs the translation invariant feature $S_0 = x * \varphi_J$ at the largest scale $(J=2)$, and there are 20 propagation operators $U_1$ in 10 directions, as follows:

$$\begin{aligned} U_1 &= \left| x * \psi_{j_1, r}(u) \right| \quad \text{with } j = 0, 1 \text{ and } r = 1, 2, \ldots, 10 \\ &= \{ \left| x * \psi_{0_1, r}(u) \right|, \left| x * \psi_{1_1, r}(u) \right| \} \quad \text{with } r = 1, 2, \ldots, 10 \end{aligned} \quad (8)$$

When $m=1$, wavelet scattering calculates 20 translation-invariant features at the largest scale using the propagation operators $U_1$ obtained from the previous layer:

$$S_1 = U_1 * \varphi_J. \tag{9}$$

Therefore, in the present work, a total of 21 translation-invariant and deformation-invariant features were obtained using the wavelet scattering transformation with the largest scale being 2 and scattering level of 1. Among these features, there are not only low-frequency components in the sense of traditional wavelet transformation, but also low-frequency components encoded in high-frequency components along multiple directions, which may provide more useful textures for image fusion.

**2.2 Interscan motion correction through inverse wavelet scattering transform using CNN**

Diffusion signals in the case of free breathing are affected by respiratory and cardiac motions, leading to phase-related artifacts that manifest as a severe signal loss in DW images. To correct interscan motions, we first apply wavelet scattering on the registered multi-TDs DW images, and then fuse the obtained multiple TDs feature maps based on a certain rule. Finally, the fused feature is mapped into a spatial DW image through inverse wavelet scattering transform using CNN.

*A. Feature maps fusion*

In order to compensate for signal loss caused by heart motion, we select 1 TD DW image as the reference, and register the other TD DW images on such reference with a rigid registration. Then, the wavelet scattering feature maps are extracted from 10 registered TDs DW images respectively. By comparing the scattering features extracted from DW images with less motion influence and those with more motion influence (Fig.3 (a)) and analyzing the corresponding histograms (Fig. 3(b)), we observe that the distribution of the scattering features of the DW image greatly affected by motions is narrower than that of the DW image less influenced by motions, which indicates that the image greatly affected by motion has smaller scattering coefficients. Based on this observation, the following scattering feature fusion rule was used. Firstly, we compared the scattering feature maps of two consecutive TDs DW images pixel by pixel and replaced the smaller pixel value by the larger pixel value to obtain a fused feature map. Then, the fused feature map was fused with the scattering feature map of the third TD DW image in the same pixel-by-pixel way. This process was repeated until all the TDs DW images are fused, thus yielding a final single fused scattering feature maps that takes advantages of the scattering features of the 10 TDs DW images and thus compensating for the motion effect to some extent.

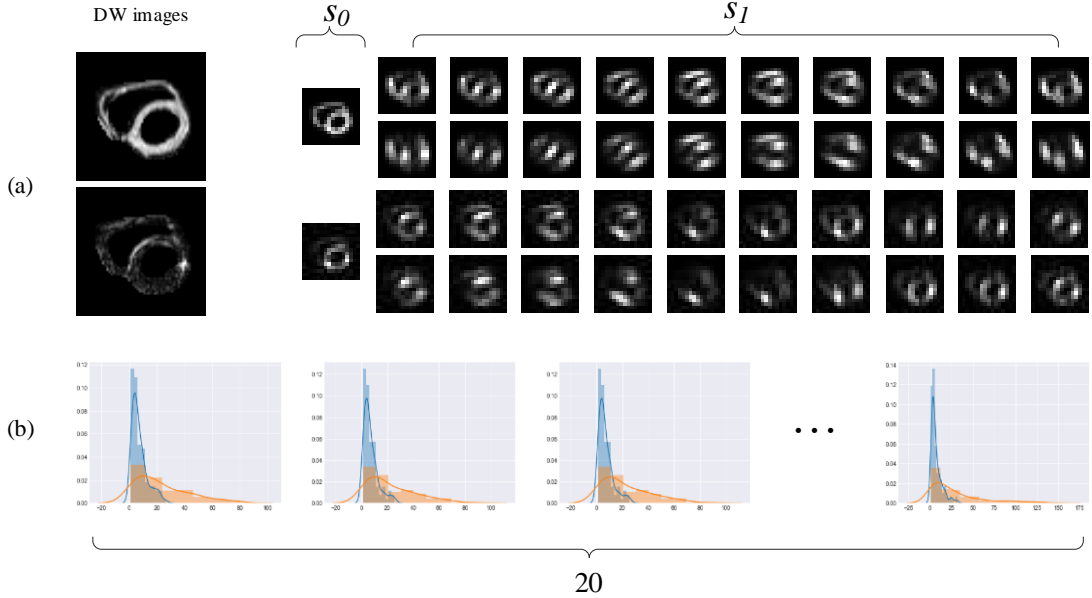

Figure 3. Variation of scattering features of DW images with different motion influences (a) and the corresponding histograms (b). In (a), the first row is the DW image with less motion influence and the corresponding scattering features where $S_0$ is the scattering feature map extracted from low-frequency component and $S_1$ are the scattering feature maps extracted from low-frequency components of high-frequency components in different directions. The second row shows the DW image with more motion influence and the corresponding scattering features. In (b) are illustrated the histograms of some scattering features $S_1$, in which the orange part represents the histograms of scattering features extracted from less motion-influenced DW images and the blue part the histograms of scattering features extracted from more motion-influenced DW images.

### B. *Inverse wavelet scattering transform using CNN*

Once we obtained the fused feature maps where motion effects are compensated for to some extent, we now need to reconstruct DW images from these fused feature maps. Since wavelet scattering transform is not exactly invertible [27], we propose to use a network based on CNN layers and residual block to achieve the inverse wavelet scattering transform by mapping multiple scattering feature maps into one single DW image. The architecture of the proposed network is depicted in Figure 4. It consists of two parts: Encoder and Decoder. The encoder-decoder structure has proven to be effective for image processing applications [28][29][30][31]. It however cannot be directly used for our present problem, which is how to reconstruct a single image from multiple feature maps. Such reconstruction requires a large receptive field. Although increasing the number of layers in the network will increase the receiving domain, the network can cause several problems: (1) A large number of parameters make the calculation amount too great, (2) More convolutional layers slow convergence speed, (3) The spatial size of the feature map generated by encoder would be too small to maintain the spatial information for reconstruction.

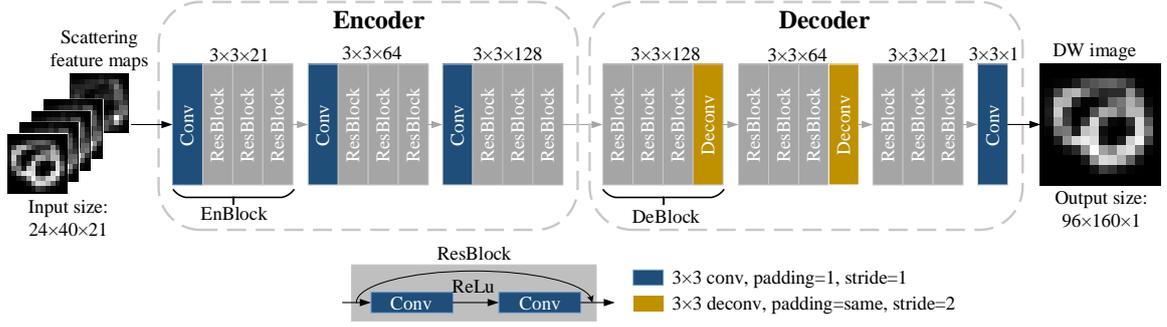

Figure 4. Architecture of the designed encoder-decoder network used to achieve inverse wavelet scattering transform.

To cope with these problems, we design the encoder-decoder network architecture as illustrated in Figure 4. The Encoder contains three Enblocks, each Enblock consisting of one convolution layer and three Resblocks for extracting deep features. The Decoder contains two Deblocks, three Resblocks and one convolution layer, each Deblock consisting of three Resblocks and one deconvolution layer. Decoder is used to double the spatial size of feature maps and reduce the number of channels. The introduction of Resblocks aims to increase the receptive field and effectively solve the problem caused by too many network layers. In training phase, the input to the network is the scattering feature maps with a size of 20×40 extracted from each of TD DW images, and the output of the network is the corresponding DW images with a size of 96×160. We implemented the proposed network using Tensorflow in python language. The $L_2$ loss was used to train the network. The used activation function was ReLu. Adam was applied to optimize the network. The learning rate, momentum, momentum2 were 0.0001, 0.9 and 0.999, respectively. In the experiments, the network converged after 260,000 iterations. Once the network was well trained, we inputted the fused scattering feature maps into the network, the output of which is our desired DW images with motion compensation.

## 2.3 Simulation of *in vivo* cardiac DTI data

To better assess the performance of the proposed WSCNN method for reducing signal loss due to motion, we first apply it to simulated *in vivo* cardiac DTI data. To make the simulated data realistic, we used the DTI data of an *ex vivo* human heart. The *ex vivo* DTI data were acquired on Siemens 1.5 T MAGNETOM Avanto in the Neuro-Cardiology Hospital of Lyon with the following settings: echo time = 69 ms, repetition time = 6500 ms, slice thickness = 2 mm, FOV = 128 × 128 mm$^2$, number of slices = 54, slice size = 128 × 128, diffusion sensitivity b = 1000 s mm$^{-2}$, and diffusion gradient directions = 12. The sequence used is a 2-D EP diffusion-weighted sequence. As illustrated in Figure 5, we use the following procedure to simulate cardiac motion from *ex-vivo* human heart:

Step 1: Select 10 slices (the same number of slices as in *in vivo* DTI case) from *ex vivo* DW images. Put these *ex vivo* DW images into a matrix of the same size (90 × 160) as that of *in vivo* DW images and use the mask to keep only the myocardium.

Step 2: Through rigid registration, the *ex vivo* DW images were mapped to the *in vivo* DW images. (the registration result obtained after this step was taken as the TD1)

Step 3: Fix one TD DW image of the *in vivo* heart and calculate the deformation field from the other 9 TD DW images with respect to this TD DW image. Apply the 9 deformation fields to the *ex* *vivo* DW image to obtain the simulated *in vivo* DW images corresponding to 9 TDs (figure 5: TD2 to TD10).

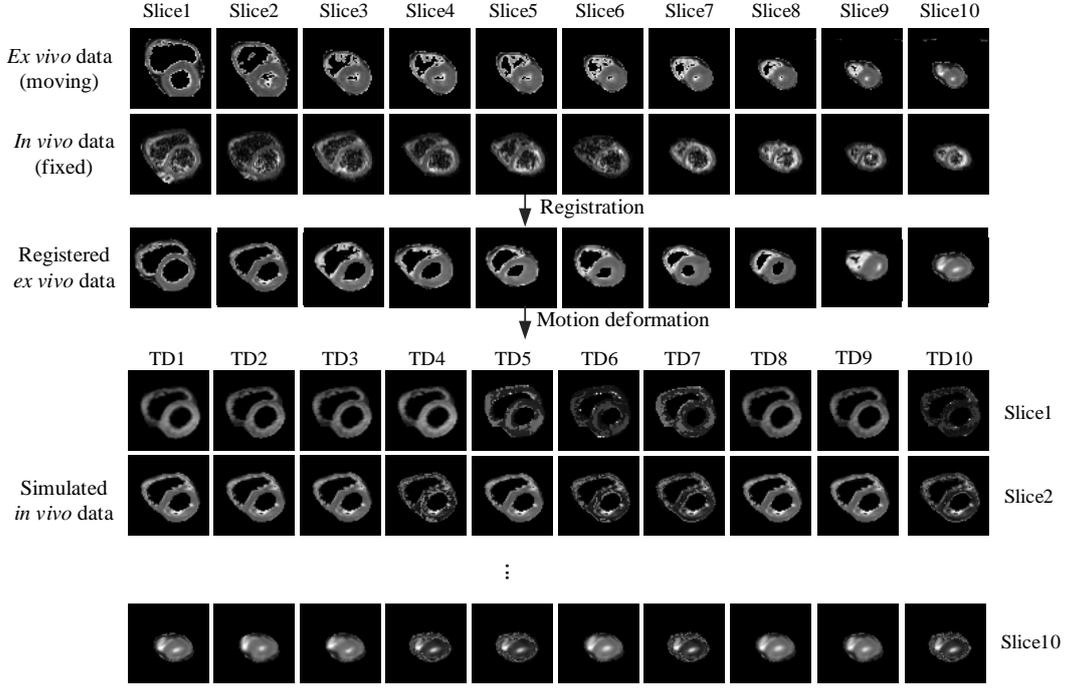

Figure 5. The process of simulating *in vivo* DW images from *ex vivo* DW images.

**2.4 Real *in vivo* cardiac DW images**

The real *in vivo* cardiac DW images come from the Neuro-Cardiology Hospital of Lyon in France. They were acquired using a multiple shifted TD acquisition strategy on a 1.5 T clinical scanner (MAGNETOM Avanto, Siemens AG, Hospital Neuro-Cardio, France). To reduce the impact of eddy current artifacts, a single-shot twice-refocused spin-echo echo-planar imaging (EPI) sequence with optimized bipolar diffusion encoding gradients was used. The gradient reversal technique was chosen to improve fat suppression. The acquisitions occurred inside an optimal time-window that corresponded to the smallest amount of cardiac motion toward the end of diastole.

For each of five healthy volunteers with age from 25 to 50 (3 males and 2 females), 10 slices (thickness=6 mm) were acquired in the short-axis view without any interslice gap. For each of the 10 slices, acquisitions in 12 diffusion gradient directions as well as for $b_0$ were performed. For each of the 12 diffusion gradient directions as well as the $b_0$, 10 shifted acquisitions with increased TD (by 10 ms) were sequentially performed, which produces 10 TDs DW images corresponding to 10 time points. Concretely, for the first time point or TD (e.g. 850 ms), a $b_0$ image and 12 DW images corresponding to the 12 diffusion gradient directions were acquired. By shifting the TD (by 10 ms), another $b_0$ image and 12 DW images (corresponded to the same 12 gradient directions) at the next time point were acquired. By repeating the TD shifting 10 times, a total of 130 multi-phase images for each cardiac layer were acquired. The total scan time was approximately 2 minutes for each cardiac layer under free-breathing conditions.

**2.5 Evaluation criteria**

To verify the performance of the proposed method for the motion compensation in *in vivo* cardiac DTI, we compared our method with existing methods in terms of cardiac fiber structures (helix angle—

HA, and transverse angle—TA) and diffusion metrics (FA and MD). HA is the angle between the short-axis plane and the projection of the fiber on the tangent plane of the epicardium, representing the longitudinal component of the fiber orientation. TA is the angle between the projection of the fiber in the short-axis plane and tangent plane, representing the transmural component of the fiber orientation, as illustrated in Figure 6.

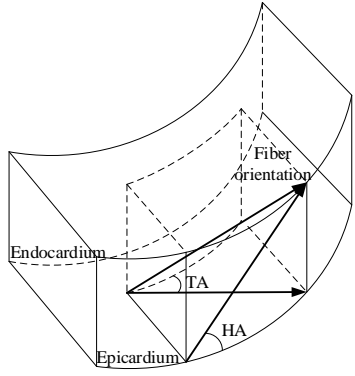

Figure 6. Helix angle (HA) and Transverse angle (TA) of a fiber.

FA is a scalar value between 0 and 1, which describes the degree of anisotropy of a diffusion process. MD reflects the overall diffusion level of molecules and the overall situation of diffusion resistance. FA and MD are defined in terms of the eigenvalues $\lambda_1$, $\lambda_2$ and $\lambda_3$ of diffusion tensor, namely

$$FA = \sqrt{\frac{1}{2}} \frac{\sqrt{(\lambda_1 - \lambda_2)^2 + (\lambda_1 - \lambda_3)^2 + (\lambda_2 - \lambda_3)^2}}{\sqrt{\lambda_1^2 + \lambda_2^2 + \lambda_3^3}}, \quad (10)$$

$$MD = \frac{(\lambda_1 + \lambda_2 + \lambda_3)}{3}. \quad (11)$$

In the simulated *in vivo* data, to quantitatively analyze the performance of motion compensations, PSNR and SSIM (between motion-compensated and reference DW images) were calculated, and the deviation angle between the fiber orientations derived from the motion-compensated DW images and the ground-truth was given. To check the integrity of 3-D fibers, we visualized myocardial 3-D fibers tracking using DSI Studio.

## 3. Results

To evaluate whether the proposed WSCNN method can more effectively compensate for signal loss caused by motion, PCATMIP, WIF and WSCNN methods were applied on both simulated and real *in vivo* DW images and the results obtained with the three methods were compared.

### 3.1 Results on simulated *in vivo* data
#### A. *Fusion of wavelet scattering feature maps*

A simulated *in vivo* DW image was decomposed into 21 feature maps using wavelet scattering, as shown in Figure 7. On the left are the simulated DW images at different time points (from TD 850 to TD 940) of the same myocardial slice and their corresponding feature maps. In the feature maps, the first one is the low-frequency component and the rest are the low-frequency components that are encoded in high-frequency components. The right side of Figure 7 is the feature maps after fusion according to the proposed fusion rules.

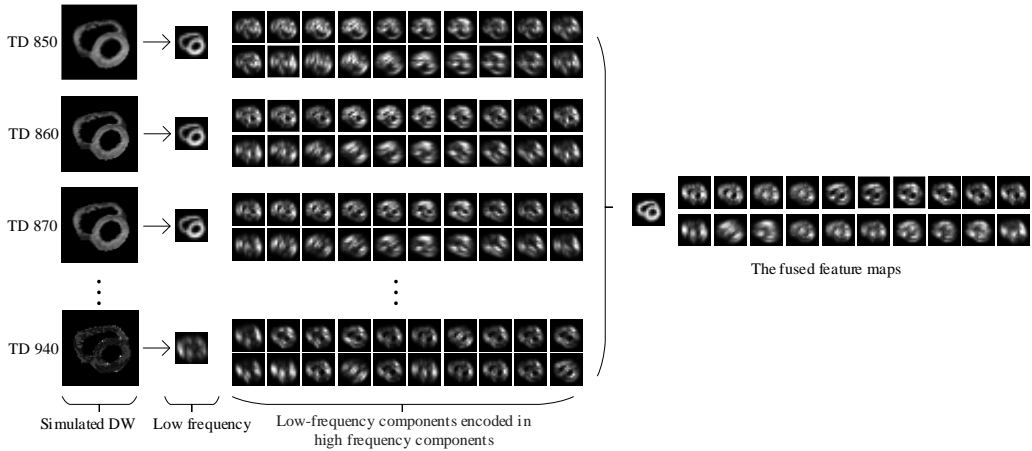

Figure 7. Simulated *in vivo* DW images (left column, from time point TD 850 to TD 940) and their feature maps obtained using wavelet scattering. The right side shows the fused feature maps.

### A. DW image fusion from fused scattering feature maps

Figure 8 shows an example of simulated *in vivo* DW images corresponding to 10 TDs with their final fused images. Figure 8(a) represents the 10 initial simulated *in vivo* DW images, in which the DW images exhibit clear signal loss due to free breathing and heart motions, especially for the images at TD 890-TD 910 and TD 940. The final images fused using PCATMIP, WIF and WSCNN are shown in Figure 8(b). The image obtained using the WSCNN method shows visually better quality than the other two methods. More quantitative assessment is given in Figure 9 in terms of PSNR and SSIM. The DW image corrected using WSCNN has higher PSNR and SSIM than those corrected using WIF and PCATMIP.

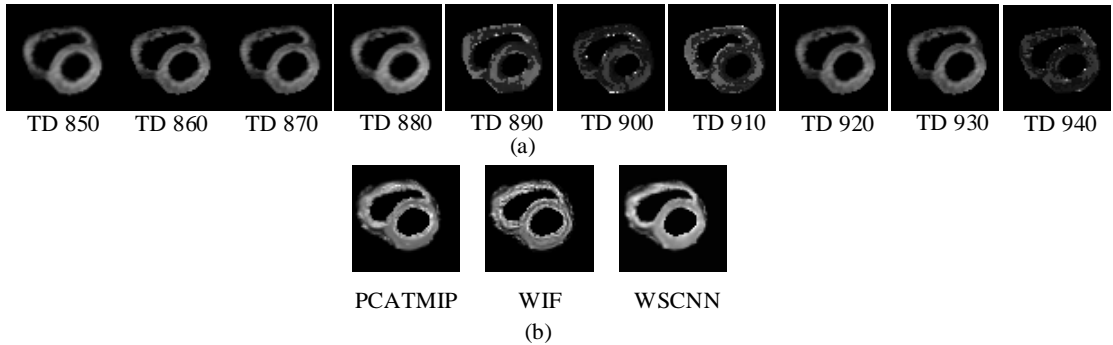

Figure 8. Fusion results on simulated *in vivo* data. (a) Simulated in vivo DW images corresponding to 10 TDs. (b) DW images fused using three different methods.

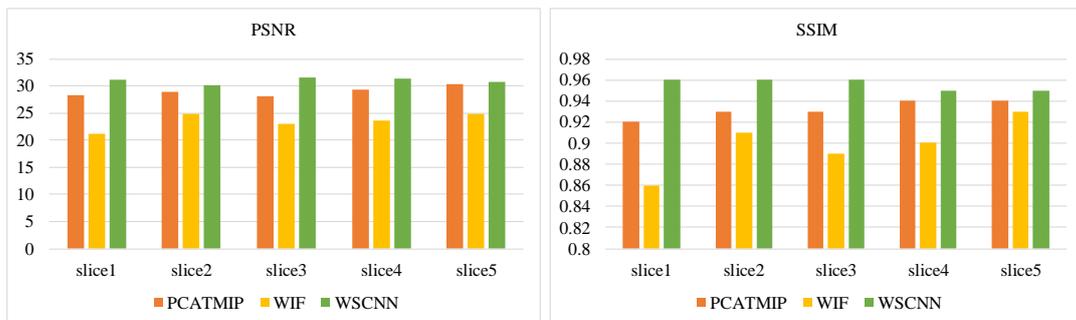

Figure 9. Quantitative assessment of simulated *in vivo* DW images corrected by different methods in terms of PSNR and SSIM.

Accurate calculation of cardiac fiber orientations and diffusion metrics from diffusion tensor imaging is very important for determining cardiac fiber structure and function. Figure 10 illustrates the fiber orientation, deviation angle between the corrected and reference fiber orientations, helix and transverse angles, FA and MD obtained with different methods. It can be seen that the fiber orientation map from the reference DW images, in which regular variation in the circular sense of fiber orientations is clearly visible. Due to the presence of motion, such regular circular variation of fiber orientations is disturbed, especially in the red-boxed areas in Figure 10(a). Although the use of PCATMIP and WIF corrected the fiber orientations to some extent, the difference between the corrected and reference fiber orientations is still lost obvious. Uncorrected DW images suffer from severe signal loss due to motion, which causes incomplete and disoriented myocardial fibers, as shown in Figure 10(b). Compared to PCATMIP and WIF, the proposed WSCNN method compensates better for motion effect in the DW images and as a result makes fiber orientations closer to the reference. This can be more readily seen in the deviation angle maps (Figure 10(c)). A smaller deviation angle implies that the corrected DW image is closer to the reference image. Compared to the corrections by PCATMIP and WIF, there are more blue parts (0° angle values) in the deviation angle map of WSCNN. From helix angles (Figure 10(d)), we observe that, compared to the reference helix angle map, the circular 0°-contour at middle myocardium is severely corrupted due to motion influence. After correction by PCATMIP or WIF, the helix angle map presents intermittently 0° values in the middle myocardium but the circular 0°-contour is still rather discontinuous. In contrast, the helix angle map from WSCNN correction is much closer to the reference, demonstrating that WSCNN corrects better motion-affected DW images. With regard to transverse angle (Figure 10(e)), most of its values are between -2° and 1° in WSCNN, between -21° and 1° in PCATMIP, and between -3° and 2° in WIF. Concerning diffusion metrics (Figure 10(f) and (g)), FA and MD maps derived from WSCNN are much closer to the reference ones with respect to those from PCATMIP and WIF.

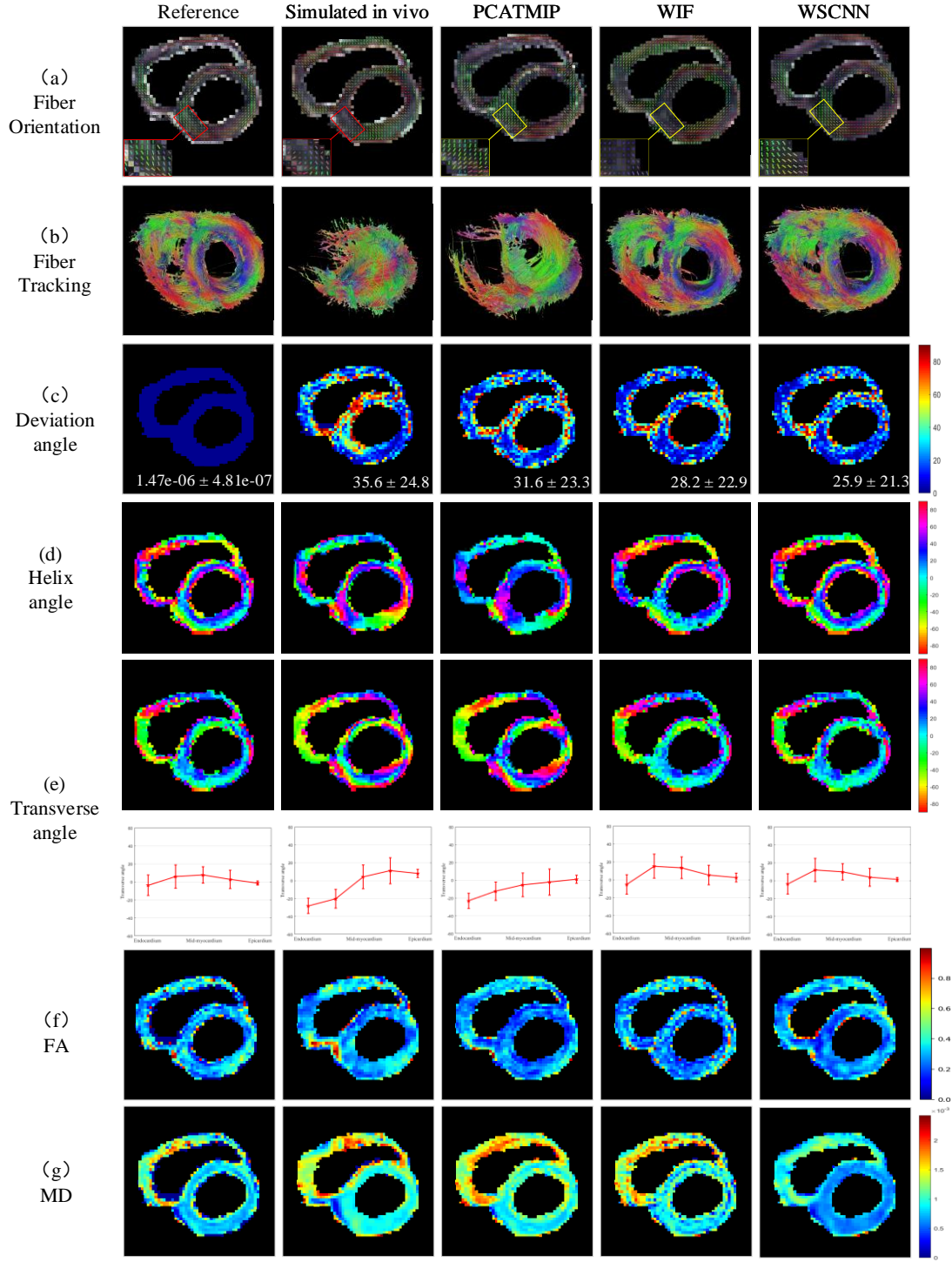

Figure 10 Fusion results obtained using different methods on simulated in vivo data. (a) Fiber orientation maps. (b) 3-D fiber tracking. (c) Maps of deviation angle between reference and corrected fiber orientations. (d) and (e) are helix and transverse angle maps. (f) and (g) are FA and MD maps. The MD values are in unit of $10^{-3}$ mm$^2$/s.

To quantitatively compare FA and MD, in Table 1 are given FA and MD values. Compared to the reference value, FA value of the simulated in vivo heart affected by motion increased by 0.07 and the MD value decreased by 0.35. This implies that cardiac motion leads to overestimated FA and underestimated MD. In Table 1. Compared to FA and MD values of the simulated in vivo heart, FA and MD values decreased respectively by 0.21 and 0.43 with PCATMIP, FA values decreased by 0.16 and MD value increased by 0.43 with WIF, FA values decreased by 0.14 and MD values increased by 0.61 with WSCNN. Clearly, FA and MD values from WSCNN is the closest to the reference ones, which demonstrates that the proposed WSCNN method can effectively correct the data affected by motion.

Table 1. FA and MD values (means ± SD) of the corrected DW image by the three methods.

|  | FA ± SD | MD ± SD |
|---|---|---|
| Reference (*ex vivo*) | 0.38 ± 0.27 | 2.12 ± 4.76 |
| Simulated *in vivo* | 0.45 ± 0.30 | 1.77 ± 4.05 |
| PCATMIP | 0.24 ± 0.18 | 1.34 ± 0.96 |
| WIF | 0.29 ± 0.20 | 2.64 ± 1.06 |
| WSCNN | 0.31 ± 0.22 | 2.38 ± 2.47 |

### 3.2 Results on real *in vivo* cardiac DW images

In the above subsection, the proposed WSCNN method has been proven to be effective in correcting simulated *in vivo* data. In the following, we further apply it to real *in vivo* data.

*A. Fusion of wavelet scattering feature maps*

The left side of Figure 11 shows real *in vivo* DW images at TDs from TD 850 to TD 940 and the corresponding feature maps obtained using wavelet scattering. The right side shows the fused feature maps.

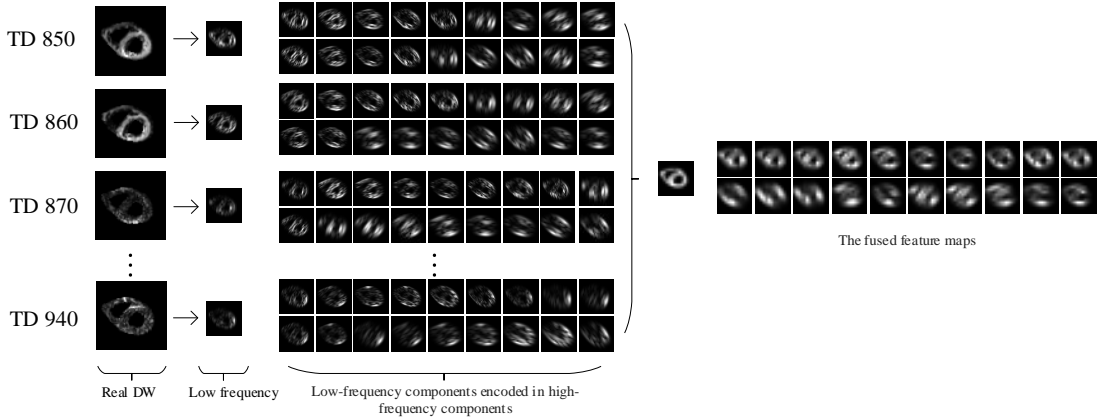

Figure 11. The left side is the wavelet scattering feature maps derived from real *in vivo* DW images at TDs from 850 ms to 940 ms, and the right side shows the fused feature maps.

*B. DW image fusion from fused scattering feature maps*

Figure 12(a) shows the free-breathing short-axis DW images from 10 repetitions acquired at different time points in one gradient diffusion direction during diastole on a volunteer. Although the DW images were acquired at the end of diastole with minimal cardiac motion, signal loss in some DW images (for example TD 910, TD 930 and TD 940) is still visible due to respiratory and heart motions. The results of compensating for the signal loss, achieved after applying PCATMIP, WIF and WSCNN on the 10 TDs DW images, are shown in the fused DW images of Figure 12(b). More quantitatively, we calculated the SNRs of DW images corrected by the different methods. The SNRs of the images corrected by PCATMIP, WIF and WSCNN are 21.3, 20.5 and 21.7 dB, respectively. We observe that the images corrected with WSCNN archives the highest SNR.

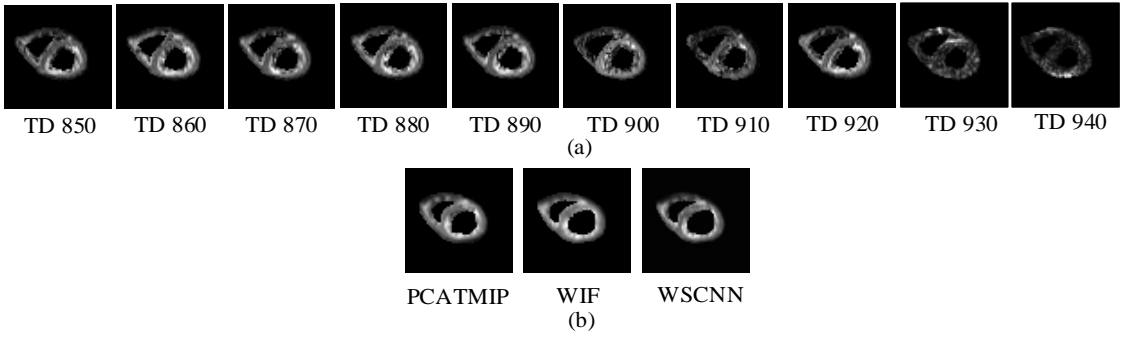

Figure 12. Ten TDs DW images acquired along one gradient diffusion direction from volunteer (a) and the DW images corrected using PCATMIP, WIF and WSCNN methods.

Figure 13 shows the fiber orientation, fiber tracking, helix and transverse angles, and FA and MD maps obtained with the three methods. We can clearly observe more circularly regular variation of fiber orientations in left ventricle after WSCNN correction with respect to the other two methods (Figure 13(a)). In the fiber tracking results (Figure 13(b)), the fiber tracks obtained with WSCNN are much better than those obtained with the other two methods, with a clear helix structure and less fibers with irregular orientations. To further assess the performance of WSCNN method with respect to the other two methods, HA and TA maps are given in Figure 13(c) and (d). We observe a clear and clean transmural variation pattern of HA after WSCNN correction, which changes from positive to negative when radially going from endocardium to epicardium, which conforms to the helix structure assumption. Such transmural HA variation is rather noisy in the PCATMIP or WIF results. Moreover, the HA range is wider in the WSCNN result (from 40° to -30°) than in the the PCATMIP (from 10° to -10°) or WIF (from 10° to -25°) result. With regard to transverse angle (Figure 13(d)), most of its values are between 7° and -5° in WSCNN, between -21°and -1° in PCATMIP, and between 10°and -17° in WIF. Concerning the diffusion metrics (Figure 13(e) and (f)), FA values obtained with WSCNN are bigger at the mid-myocardium than at epicardium and endocardium, which confirms the findings of Ariga et al. [32]. Such characteristic was not found in FA maps obtained with PCATMIP or WIF. The MD map obtained with WSCNN is much clearer than those obtained with PCATMIP or WIF.

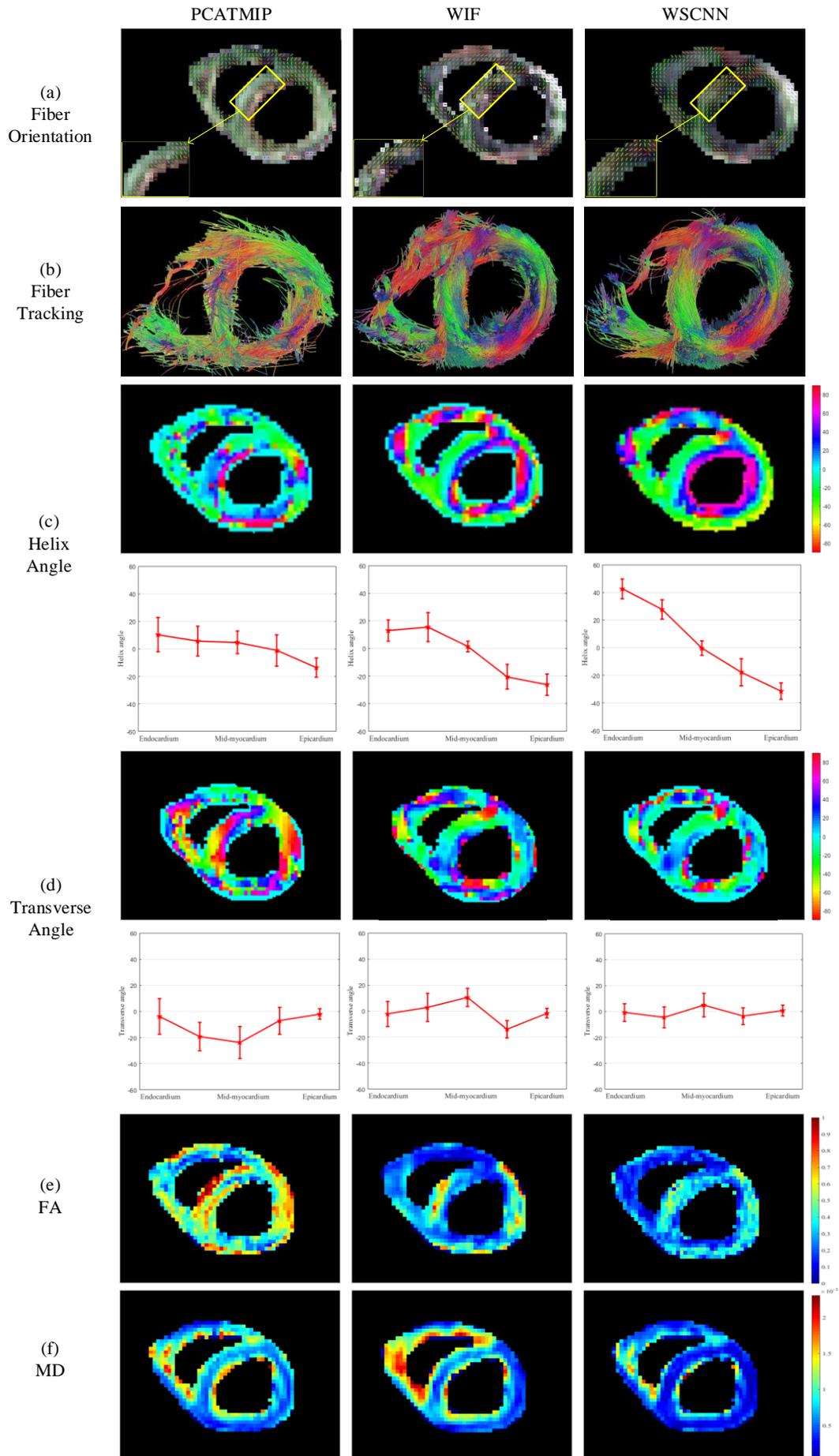

Figure 13. Fiber orientation, fiber tracking, helix and transverse angle, FA, and MD maps obtained from acquired in vivo DW images corrected using PCATMIP, WIF and WSCNN. (a) Fiber orientation maps. (b) 3-D fiber tracking. (c) helix angle maps and plots of the helix angle with respect to the transmural positions from the endocardium to the epicardium (d) transverse angle maps. (e) and (f) are FA and MD maps. The MD values are in unit of $10^{-3}$ mm$^2$/s.

To further assess the performance of the proposed method, the average HA values of different AHA segments of five volunteers obtained with the three methods are presented as bulls-eye plots in Figure 14, in which the myocardium was divided into three concentric annuluses from endocardium to epicardium, which correspond to the endocardium, mid-myocardium and epicardium. In addition, six AHA segments were divided, which correspond respectively to anterior, anteroseptal, inferoseptal inferior, inferolateral and anterolateral zones.

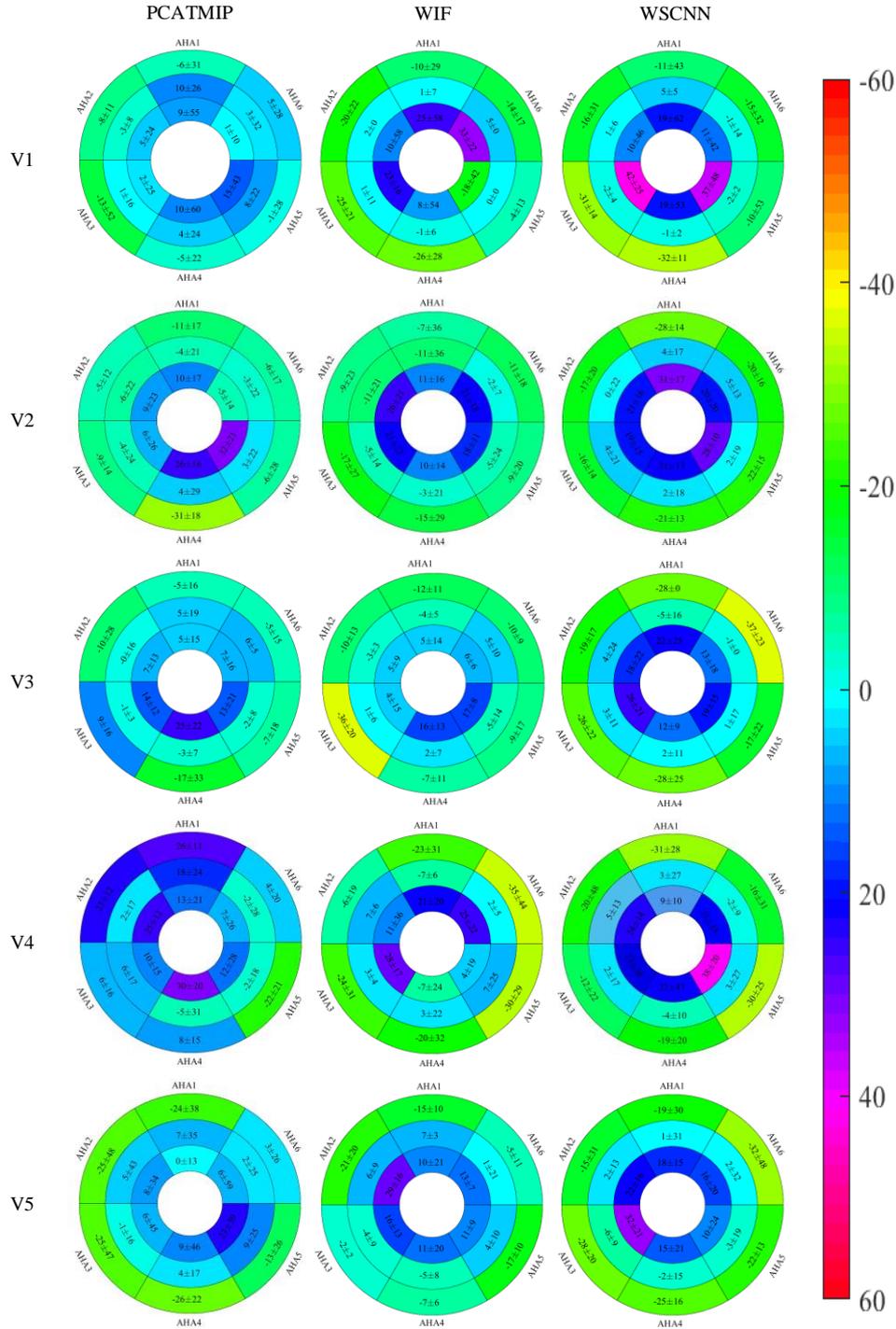

Figure 14. Bulls-eye plots of helix angles in six AHA segments with PCATMIP, WIF and WSCNN methods for different volunteers. AHA 1 to AHA 6 correspond respectively to anterior, anterolateral, inferolateral, inferior, inferoseptal and anteroseptal zones.

It can be readily seen that, in most of the AHA segments, the mean HA values obtained from PCATMIP are not in line with the property that varying from positive at endocardium to negative at epicardium, especially in the epicardium part where PCATMIP exhibits a larger positive value. With WIF, the mean HA in the endocardium is negative for V1 (AHA5) and V4 (AHA4), and the variation range of helix angles from endocardium to epicardium is small. That does not conform to the variation property of helix angle. In contrast, the mean HA values obtained with the proposed WSCNN method have smoother and more regular variations in all the AHA segments for all volunteers.

In addition, the average FA and MD values over the left ventricles of the 5 volunteers obtained with different methods are given in Table 2. Since there is no ground-truth for real FA and MD of *in vivo* hearts, we chose to compare the here calculated FA and MD values with those obtained with advanced motion-compensation acquisition techniques. We observe that FA ($0.44 \pm 0.06$) and MD ($1.70 \pm 1.03 \times 10^{-3}$ mm2/s) values after correction by WSCNN are the closest to those obtained using the advanced acquisition techniques.

Table 2. Comparing FA and MD values obtained with different methods with the existing studies. The unit of MD is $10^{-3}$ mm$^2$/s.

|    | Stoeck et al. [21] | Nielles-Vallespin et al. [20] | Scott et al. [34] | PCA | WIF | WSCNN |
|----|---|---|---|---|---|---|
| FA | $0.61 \pm 0.04$ | $0.60 \pm 0.03$ | $0.61 \pm 0.03$ | $0.36 \pm 0.04$ | $0.38 \pm 0.06$ | $0.44 \pm 0.06$ |
| MD | $0.87 \pm 1.00$ | $0.90 \pm 0.20$ | $1.10 \pm 0.21$ | $2.50 \pm 1.29$ | $2.10 \pm 1.57$ | $1.70 \pm 1.03$ |

## 4. Discussion

Investigating the architecture of *in vivo* hearts is still a challenge due to motion effects. In the present work, we have proposed to use wavelet scattering features and CNN-based model (WSCNN) to compensate for physiological motion in *in vivo* cardiac DTI with free breathing. We demonstrated that, on the simulated *in vivo* data, the proposed WSCNN method was able to effectively compensate for the motion and consequently generate the fused DW images of high quality as well as coherent fiber structure closer to the referenced one. On acquired *in vivo* data that was greatly influenced by noise, WSCNN still allowed us to clearly observe the helical structure of the myocardium with a circumferential zero contour at mid-myocardium and a smooth positive-negative transmural transition from endocardium to epicardium. This is consistent with previous study of *ex vivo* hearts [33].

To the best of our knowledge, it is the first time that transformation invariant features and deep learning model were combined to deal with motion-induced problems in *in vivo* DTI. Although there exist a few acquisition techniques aimed at reducing the effects of bulk motion for in vivo cardiac DTI, such as bipolar diffusion encoding gradient pulses [14], simulated echoes over two cardiac cycles or a single short acquisition with navigator-based gating [15], and diffusion encoding schemes that compensate for the first-, second- or high-order motion [14][11][10], our method presents the particularity of using clinically available sequences to acquire in vivo data in combination with image postprocessing techniques to cope with motion effects. Compared to existing postprocessing methods of compensating for motions in free breathing cardiac DTI acquisitions, such as PCATMIP [25] and WIF [18], the proposed WSCNN method worked on translation-invariant wavelet scattering features extracted from DW images in multiple diffusion directions, which enabled us to make use of more accurate information to fuse different TDs DW images. That explains why motion correction was much better with our method than with PCAIMIP or WIF.

Since wavelet scattering transform itself is non-invertible, to reconstruct the fused DW images from the fused scattering features, we used a CNN-based network to learn the relationship between the wavelet scattering features and the corresponding DW images. In the present work, we acquired 10 slices along 12 diffusion gradient directions at 10 different trigger delay moments for each heart, which represents, if the $B_0$ images are also counted, 1300 samples for training. Such a dataset size is enough to make the network converge. However, if the acquisition protocol available allows acquiring only one slice along a few diffusion directions without trigger delay or repetition acquisitions, the samples will not be sufficient for training. This is a limitation of our method.

Bulk motion also influences (spatial) phase while, in our present study, motion compensation was implemented only on magnitude images. Performing motion compensation on complex data (if available with advanced sequences) containing both phase and magnitude would constitute an interesting future work. On the other hand, it takes a longer time to scan DW images along multiple diffusion gradient directions at multiple trigger delay moments. For instance, the average acquisition time for one volunteer is about 25 minutes, during which only 10 slices are acquired, which represents a too large slice interval to get an accurate description of myocardium structure of a whole heart. In the future, it would be interesting to develop postprocessing methods allowing for augmenting dataset through obtaining more slices, which could enable us to recover possible missing fiber structure details and compensate for motion effect at the same time.

It should be noted that, although the current work focused on cardiac DTI, the proposed method can be applied to other diffusion MRI of any other organ to tackle the problem of motion effects.

## 5. Conclusion

We have proposed a novel method called WSCNN to investigate *in vivo* myocardium fiber structures in DTI with free-breathing acquisitions. The method consists of extracting and fusing wavelet scattering features of DW images acquired at different TDs, and reconstructing motion-compensated DW images using a CNN-based invertible wavelet scattering. The results on both simulated and acquired in vivo cardiac DW images showed that the proposed WSCNN method effectively compensates for motion-induced signal loss and produces *in vivo* cardiac DW images with better quality and more coherent fiber structures with respect to existing PCATMIP and WIF methods, which makes it an interesting method for measuring correctly diffusion properties of the in vivo human heart in DTI under free breathing.

**References**

[1] G. A. Roth *et al.*, "Demographic and Epidemiologic Drivers of Global Cardiovascular Mortality," *N. Engl. J. Med.*, vol. 372, no. 14, pp. 1333–1341, 2015.

[2] E. R. R. *et al.*, "In vivo measurement of water diffusion in the human heart," *Magn. Reson. Med.*, vol. 32, no. 3, pp. 423–428, 2018.

[3] C. Frindel, M. Robini, J. Schaerer, P. Croisille, and Y. M. Zhu, "A graph-based approach for automatic cardiac tractography," *Magn. Reson. Med.*, vol. 64, no. 4, pp. 1215–1229, 2010.

[4] C. Frindel, M. Robini, P. Croisille, and Y. M. Zhu, "Comparison of regularization methods for human cardiac diffusion tensor MRI," *Med. Image Anal.*, vol. 13, no. 3, pp. 405–418, 2009.

[5] M. T. Wu *et al.*, "Diffusion tensor magnetic resonance imaging mapping the fiber architecture remodeling in human myocardium after infarction: Correlation with viability and wall motion," *Circulation*, vol. 114, no. 10, pp. 1036–1045, 2006.

[6] E. X. Wu *et al.*, "MR diffusion tensor imaging study of postinfarct myocardium structural remodeling in a porcine model," *Magn. Reson. Med.*, vol. 58, no. 4, pp. 687–695, 2007.

[7] F. Yang, Y. M. Zhu, J. H. Luo, M. Robini, J. Liu, and P. Croisille, "A comparative study of different level interpolations for improving spatial resolution in diffusion tensor imaging," *IEEE J. Biomed. Heal. Informatics*, vol. 18, no. 4, pp. 1317–1327, 2014.

[8]  P. F. Ferreira *et al.*, "In vivo cardiovascular magnetic resonance diffusion tensor imaging shows evidence of abnormal myocardial laminar orientations and mobility in hypertrophic cardiomyopathy," *J. Cardiovasc. Magn. Reson.*, vol. 16, no. 1, pp. 1–16, 2014.

[9]  C. Nguyen *et al.*, "In vivo three-dimensional high resolution cardiac diffusion-weighted MRI: A motion compensated diffusion-prepared balanced steady-state free precession approach," *Magn. Reson. Med.*, vol. 72, no. 5, pp. 1257–1267, 2014.

[10] H. E. Welsh CL, DiBella EV, "Higher-Order Motion-Compensation for In Vivo Cardiac Diffusion Tensor Imaging in Rats," *IEEE Trans. Med. Imaging*, vol. 34, no. 9, pp. 1843–1853, 2015.

[11] S. Christian, C. von Deuster, G. Martin, A. David, and K. Sebastian, "Second-order motion-compensated spin echo diffusion tensor imaging of the human heart," *Magn. Reson. Med.*, vol. 75, no. 4, pp. 1669–1676, 2015.

[12] C. Mekkaoui *et al.*, "Myocardial infarct delineation in vivo using diffusion tensor MRI and the tractographic propagation angle," *J. Cardiovasc. Magn. Reson.*, vol. 15, no. S1, pp. 2–4, 2013.

[13] D. E. Sosnovik *et al.*, "Microstructural impact of ischemia and bone marrow-derived cell therapy revealed with diffusion tensor magnetic resonance imaging tractography of the heart in vivo," *Circulation*, vol. 129, no. 17, pp. 1731–1741, 2014.

[14] U. Gamper, P. Boesiger, and S. Kozerke, "Diffusion imaging of the in vivo heart using spin echoes-considerations on bulk motion sensitivity," *Magn. Reson. Med.*, 2007.

[15] S. Nielles-Vallespin *et al.*, "In vivo diffusion tensor MRI of the human heart: Reproducibility of breath-hold and navigator-based approaches," *Magn. Reson. Med.*, 2013.

[16] A. Z. Lau, E. M. Tunnicliffe, R. Frost, P. J. Koopmans, D. J. Tyler, and M. D. Robson, "Accelerated human cardiac diffusion tensor imaging using simultaneous multislice imaging," *Magn. Reson. Med.*, 2015.

[17] D. B.M.A. *et al.*, "In Vivo cardiac diffusion-weighted magnetic resonance imaging: Quantification of normal perfusion and diffusion coefficients with intravoxel incoherent motion imaging," *Invest. Radiol.*, 2012.

[18] H. Wei *et al.*, "Free-breathing diffusion tensor imaging and tractography of the human heart in healthy volunteers using wavelet-based image fusion," *IEEE Trans. Med. Imaging*, vol. 34, no. 1, pp. 306–316, 2015.

[19] M. Kevin *et al.*, "In vivo free-breathing DTI and IVIM of the whole human heart using a real-time slice-followed SE-EPI navigator-based sequence: A reproducibility study in healthy volunteers," *Magn. Reson. Med.*, vol. 76, no. 1, pp. 70–82, 2015.

[20] S. Nielles-Vallespin *et al.*, "In vivo diffusion tensor MRI of the human heart: Reproducibility of breath-hold and navigator-based approaches," *Magn. Reson. Med.*, vol. 70, no. 2, pp. 454–465, 2013.

[21] C. T. Stoeck *et al.*, "Dual-phase cardiac diffusion tensor imaging with strain correction," *PLoS One*, vol. 9, no. 9, pp. 1–12, 2014.

[22] E. M. Tunnicliffe *et al.*, "Intercentre reproducibility of cardiac apparent diffusion coefficient and fractional anisotropy in healthy volunteers," *J. Cardiovasc. Magn. Reson.*, vol. 16, no. 1, pp. 1–12, 2014.

[23] M. Froeling, G. J. Strijkers, A. J. Nederveen, and P. R. Luijten, "Whole heart DTI using asymmetric bipolar diffusion gradients," *J. Cardiovasc. Magn. Reson.*, vol. 17, no. S1, pp. 1–2, 2015.

[24] W. Y. I. Tseng, T. G. Reese, R. M. Weisskoff, and V. J. Wedeen, "Cardiac diffusion tensor MRI in vivo without strain correction," *Magn. Reson. Med.*, vol. 42, no. 2, pp. 393–403, 1999.

[25] V. M. Pai, S. Rapacchi, P. Kellman, P. Croisille, and H. Wen, "PCATMIP: Enhancing signal intensity in diffusion-weighted magnetic resonance imaging," *Magn. Reson. Med.*, vol. 65, no. 6, pp. 1611–1619, 2011.

[26] S. Mallat, "Group Invariant Scattering," *Commun. Pure Appl. Math.*, 2012.

[27] J. Bruna and S. Mallat, "Invariant scattering convolution networks," *IEEE Trans. Pattern Anal. Mach. Intell.*, vol. 35, no. 8, pp. 1872–1886, 2013.

[28] Z. Liu, R. A. Yeh, X. Tang, Y. Liu, and A. Agarwala, "Video Frame Synthesis Using Deep Voxel Flow," in *Proceedings of the IEEE International Conference on Computer Vision*, 2017.

[29] S. Su, M. Delbracio, J. Wang, G. Sapiro, W. Heidrich, and O. Wang, "Deep video deblurring for hand-held cameras," in *Proceedings - 30th IEEE Conference on Computer Vision and Pattern Recognition, CVPR 2017*, 2017.

[30] X. Tao, H. Gao, R. Liao, J. Wang, and J. Jia, "Detail-Revealing Deep Video Super-Resolution," in *Proceedings of the IEEE International Conference on Computer Vision*, 2017.

[31] N. Xu, B. Price, S. Cohen, and T. Huang, "Deep image matting," in *Proceedings - 30th IEEE Conference on Computer Vision and Pattern Recognition, CVPR 2017*, 2017.

[32] R. Ariga *et al.*, "Identification of Myocardial Disarray in Patients With Hypertrophic Cardiomyopathy and Ventricular Arrhythmias," *J. Am. Coll. Cardiol.*, vol. 73, no. 20, pp. 2493–2502, 2019.

[33] H. Lombaert *et al.*, "Human atlas of the cardiac fiber architecture: Study on a healthy population," *IEEE Trans. Med. Imaging*, vol. 31, no. 7, pp. 1436–1447, 2012.

[34] A. D. Scott *et al.*, "An in-vivo comparison of stimulated-echo and motion compensated spin-echo sequences for 3 T diffusion tensor cardiovascular magnetic resonance at multiple cardiac phases," *J. Cardiovasc. Magn. Reson.*, vol. 20, no. 1, pp. 1–15, 2018.